\documentclass[twocolumn,amsmath,amssymb,floatfix]{revtex4}

\usepackage{graphicx}
\usepackage{dcolumn}
\usepackage{bm}

\graphicspath{{Figures/}}
\setkeys{Gin}{width=\linewidth}

\begin{document}

\title{Aharonov-Bohm oscillations in the presence of strong spin-orbit interactions}

\author{Boris Grbi\'{c}$^{*}$, Renaud Leturcq$^{*}$, Thomas Ihn$^{*}$,
Klaus Ensslin$^{*}$, Dirk Reuter $^{+}$, and Andreas D.
Wieck$^{+}$}

\affiliation{$^{*}$Solid State Physics Laboratory, ETH Zurich,
  8093 Zurich, Switzerland \\$^{+}$Angewandte Festk\"{o}rperphysik,
Ruhr-Universit\"{a}t Bochum, 44780 Bochum, Germany}

\begin{abstract}

We have measured highly visible Aharonov-Bohm (AB) oscillations in
a ring structure defined by local anodic oxidation on a p-type
GaAs heterostructure with strong spin-orbit interactions. Clear
beating patterns observed in the raw data can be interpreted in
terms of a spin geometric phase. Besides $h/e$ oscillations, we
resolve the contributions from the second harmonic of AB
oscillations and also find a beating in these $h/2e$ oscillations.
A resistance minimum at $B=0$ T, present in all gate
configurations, is the signature of destructive interference of
the spins propagating along time-reversed paths.

\end{abstract}

\maketitle

Interference phenomena with particles have challenged physicists
since the foundation of quantum mechanics. A charged particle
traversing a ring-like mesoscopic structure in the presence of an
external magnetic flux $\Phi$ acquires a quantum mechanical phase.
The interference phenomenon based on this phase is known as the
Aharonov-Bohm (AB) effect \cite{Aharonov59}, and manifests itself
in oscillations of the resistance of the mesoscopic ring with a
period of  $\Phi_0=h/e$, where $\Phi_0$ is the flux quantum. The
Aharonov-Bohm phase was later recognized as a special case of the
geometric phase \cite{Berry84, Aharonov87} acquired by the orbital
wave function of a charged particle encircling a magnetic flux
line.

The particle's spin can acquire an additional geometric phase in
systems with spin-orbit interactions (SOI)
\cite{Loss99,Engel00,Meir89}. The investigation of this spin-orbit
(SO) induced phase in a solid-state environment is currently the
subject of intensive experimental work \cite{Morpugo98, Yau02,
Yang04, Konig06, Bergsten06, Habib07}. The common point of these
experiments is the investigation of electronic transport in
ring-like structures defined on two-dimensional (2D)
semiconducting systems with strong SOI. Electrons in InAs were
investigated in a ring sample with time dependent fluctuations
\cite{Morpugo98}, as well as in a ring side coupled to a wire
\cite{Yang04}. An experiment on holes in GaAs \cite{Yau02} showed
B-periodic oscillations with a relative amplitude $\Delta R/R <
10^{-3}$. These observations \cite{Morpugo98, Yau02} were analyzed
with Fourier transforms and interpreted as a manifestation of
Berry's phase. Further studies on electrons in a HgTe ring
\cite{Konig06} and in an InGaAs ring network \cite{Bergsten06}
were discussed in the framework of the Aharonov-Casher effect.

In systems with strong SOI, an inhomogeneous, momentum dependent
intrinsic magnetic field $\mathbf{B_{int}}$, perpendicular to the
particle's momentum, is present in the reference frame of the
moving carrier \cite{Winkler03}. The total magnetic field seen by
the carrier is therefore $\mathbf{B_{tot}}=\mathbf{B_{ext}}+
\mathbf{B_{int}}$, where $\mathbf{B_{ext}}$ is the external
magnetic field perpendicular to the 2D system and
$\mathbf{B_{int}}$ is the intrinsic magnetic field in the plane of
the 2D system present in the moving reference frame (right inset
Fig. 1(a)). The particle's spin precesses around
$\mathbf{B_{tot}}$ and accumulates an additional geometric phase
upon cyclic evolution.

Effects of the geometric phases are most prominently expressed in
the adiabatic limit, when the precession frequency of the spin
around the local field $\mathbf{B_{tot}}$ is much faster than the
orbital frequency of the charged particle carrying the spin
\cite{Loss99}. In this limit the ring can be considered to consist
of two uncoupled types of carriers with opposite spins
\cite{Stern92}. The total accumulated phase, composed of the AB
phase and the SO induced geometric phase, is different for the two
spin species, $\phi_{tot}= \phi_{AB} \pm \phi_{SO}$, and the
magnetoresistance of the ring is obtained as the superposition of
the oscillatory contributions from the two spin species. Such a
superposition is predicted to produce complex, beating-like
magnetoresistance oscillations with nodes developing at particular
values of the external B-field, where the oscillations from the
two spin-species have opposite phases \cite{Engel00}. Both $h/e$
and $h/2e$ peaks in the Fourier spectrum of the magnetoresistance
oscillations are predicted to be split in the presence of strong
SOI \cite{Malshukov99, Engel00}.

The interpretation of the split Fourier signal in Ref.
\cite{Morpugo98} has been challenged \cite{Raedt99}. The data on
p-GaAs rings \cite{Yau02} stirred an intense discussion
\cite{Chao03}. Our raw data directly displays a beating of the
$h/e$ Aharonov-Bohm oscillations. No Fourier transform is required
to verify this effect. As additional evidence we directly measure
a beating of the $h/2e$ oscillations and a pronounced and
persistent zero field magnetoresistance minimum due to destructive
interference of time-reversed paths.


The sample was fabricated by atomic force microscope (AFM)
oxidation lithography on a p-type carbon doped (100) GaAs
heterostructure, with a shallow two-dimensional hole gas (2DHG)
located 45 nm below the surface \cite{Grbic05}. An AFM micrograph
of the ring structure is shown in the inset of Fig. 1(a). The
average radius of the circular orbit is 420 nm, and the
lithographic width of the arms is 190 nm, corresponding to
an electronic width of $60-70$ nm. 
The hole density in an unpatterned sample is 3.8$\times$10$^{11}$
cm$^{-2}$ and the mobility is 200 000 cm$^2$/Vs at a temperature
of 60 mK. Therefore the Fermi wavelength is about 40 nm, and the
mean free path is 2 $\mu$m. Since the circumference of the ring is
around 2.5 $\mu$m, the transport through the ring is
quasiballistic. From the temperature dependence of the AB
oscillations we extract the phase coherence length of the holes to
be $L_{\varphi}=2$ $\mu$m at a base temperature of $T=60$ mK.

The presence of strong spin-orbit interactions in the
heterostructure is demonstrated by a simultaneous observation of
the beating in Shubnikov-de Haas (SdH) oscillations and a weak
anti-localization dip in the measured magnetoresistance of the
Hall bar fabricated on the same wafer \cite{GrbicThesis07}. In
p-type GaAs heterostuctures, Rashba SOI is typically dominant over
the Dresselhaus SOI \cite{Lu98}. The densities N$_1$=
1.35$\times$10$^{11}$ cm$^{-2}$ and N$_2$= 2.45$\times$10$^{11}$
cm$^{-2}$ of the spin-split subbands, deduced from SdH
oscillations, allow us to estimate the strength of the Rashba
spin-orbit interaction $\Delta_{SO}\approx0.8$ meV assuming a
cubic wave-vector dependence \cite{Winkler03}. Due to the large
effective mass of the holes, the Fermi energy in the system,
$E_F=2.5$ meV, is much smaller than that in electron systems with
the same density. The large ratio $\Delta _{SO}/E_F\approx 30\%$
documents the presence of strong SOI.

We have measured the four-terminal resistance of the ring in a
$^3$He/$^4$He dilution refrigerator at a base temperature of about
60 mK with lock-in techniques. A low ac current of 2 nA and 31 Hz
frequency was applied in order to prevent sample heating.

\begin{figure}
  \begin{center}
    \includegraphics{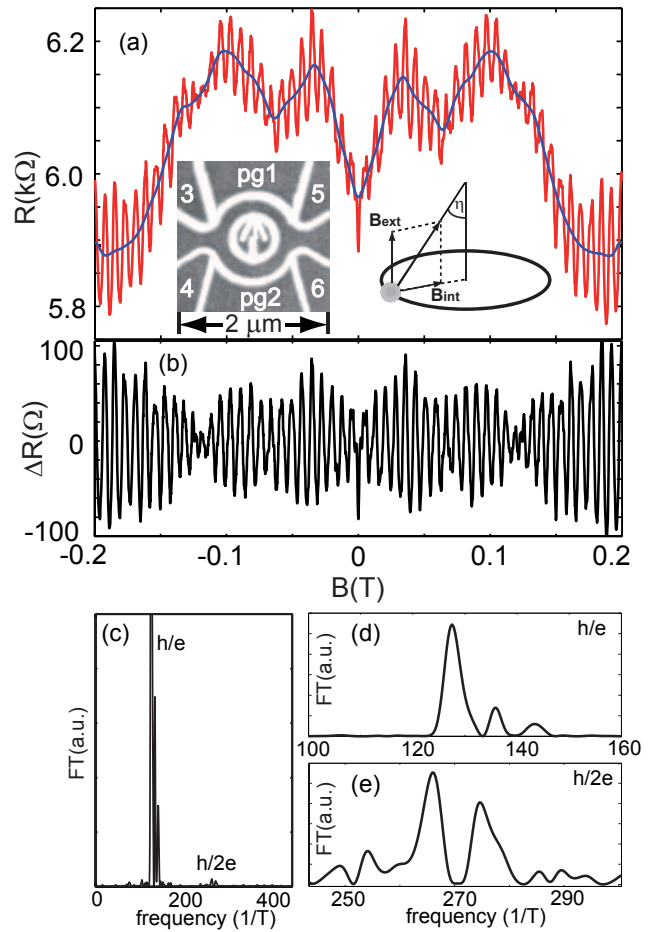}
    \caption{(color online) (a) Measured magnetoresistance of the ring (strongly
oscillating curve, red online) together with the low-frequency
background resistance (smooth curve, blue online); Left inset: AFM
micrograph of the ring with designations of the in-plane gates.
Bright oxide lines fabricated by AFM oxidation lithography lead to
insulating barriers in the 2DHG. Right inset: Scheme of a carrier
travelling around the ring in the presence of the external field
$\mathbf{B_{ext}}$ and SO induced intrinsic field
$\mathbf{B_{int}}$. (b) AB oscillations obtained after subtraction
of the low-frequency background from the raw data. A clear beating
pattern is revealed in the AB oscillations. (c) Fourier transform
spectra of the AB oscillations, revealing $h/e$ and $h/2e$ peaks.
(d) Splitting of the $h/e$ Fourier peak. (e) Splitting of the
$h/2e$ Fourier peak.}
    \label{fig1}
  \end{center}
\end{figure}

Fig. 1(a) shows the magnetoresistance of the ring (fast
oscillating curve, red online). The low-frequency background
resistance is indicated by a smooth curve (blue online). The
observed Aharonov-Bohm (AB) oscillations with a period of 7.7 mT
(frequency 130 T$^{-1}$) correspond to a radius of the holes'
orbit of 415 nm, in excellent agreement with the lithographic size
of the ring. The peak-to-peak amplitude of $\sim 200$ $\Omega$ on
a background of about 6 k$\Omega$ corresponds to a visibility
larger than $3\%$. We restrict the measurements of the AB
oscillations to magnetic fields in the range from -0.2 T to 0.2 T
in order to prevent their mixing with SdH oscillations, which
start to develop above 0.2 T. Throughout all measurements quantum
point contact gates 3,4,5 and 6 are kept at the same values.
Plunger gates pg1 and pg2 are set to V$_{pg1}=-145$ mV and
V$_{pg2}=-95$ mV in the measurements presented in Fig. 1(a).

After subtracting the low-frequency background from the raw data,
a clear beating pattern is revealed in the AB oscillations with a
well defined node at $\sim115$ mT [Fig. 1(b)], where a phase jump
of $\pi$ occurs [arrow in Fig. 2(c)]. The position of the beating
node indicates the presence of two oscillation frequencies
differing by $1/0.115 \approx 9$ T$^{-1}$.

The Fourier spectrum of the AB oscillations, taken in the
symmetric magnetic field range (-0.2 T, 0.2 T), reveals an $h/e$
peak around 130 T$^{-1}$ [Fig. 1(c)]. Zooming in on the $h/e$
peak, a splitting into 3 peaks at the frequencies 127 T$^{-1}$,
136 T$^{-1}$ and 143 T$^{-1}$ is seen. We have carefully checked
that this splitting is genuine to the experimental data and not a
result of the finite data range, by reproducing it with different
window functions for the Fourier transform. The differences of the
oscillation frequencies agree with that anticipated from the
position of the beating node in the raw data.

In contrast to the $h/e$-periodic AB oscillations, which are very
sensitive to phase changes in the ring arms,
Altshuler-Aronov-Spivak (AAS) $h/2e$ oscillations, originating
from the interference of time reversed paths, are expected to be
more robust if the microscopic configuration of the arms is
changed. Besides, $h/2e$ oscillations are less susceptible to the
details how the spin rotates when it enters the ring than $h/e$
oscillations. This is due to the fact that the geometric phase
accumulated along the paths contributing to the $h/2e$
oscillations is larger than that in the case of the $h/e$
oscillations and cannot be completely cancelled by the spin
rotations in the contacts, as in the latter case \cite{Yang04}. In
Fig. 1(c) we can identify the peak at about 270 T$^{-1}$ in the
Fourier spectrum, corresponding to $h/2e$ oscillations. If we zoom
in on it [Fig. 1(e)], we see a splitting with the two main peaks
having a separation of about 8 T$^{-1}$, similar to the $h/e$ peak
splitting. The splitting of the $h/2e$ Fourier peak arises due to
the frequency shift of the main peak by $\pm 1/B_{int}$
\cite{Engel00}, and the obtained splitting of 8 T$^{-1}$ allows to
estimate the SO induced intrinsic field to be $B_{int}\approx
0.25$T.

We now focus directly on the magnetic field-dependent resistance.
In Fig. 2(a) we present the raw data after subtracting the
low-frequency background (full line, red online) together with the
filtered $h/e$ oscillations (dashed line) \cite{Sigrist04}. The
$h/e$ contribution to the signal is the inverse Fourier transform
of the $h/e$ peak in the Fourier spectrum. We will use the
following notation below: $R_d$ denotes the raw data, $R_b$ is the
low-frequency background, $R_{h/e}$ is the inverse Fourier
transform of the $h/e$ peak and $R_{h/2e}$ is the inverse Fourier
transform of the $h/2e$ peak in the Fourier spectrum. One can see
in Fig. 2(a) that $R_d-R_b$ contains additional resistance
modulations, beyond the $h/e$ oscillations. In order to
demonstrate that those additional features are due to $h/2e$
oscillations we plot in Fig. 2(b) the difference $R_d-R_b-R_{h/e}$
(full line, red online) and the curve $R_{h/2e}$ obtained by
inverse Fourier transform of the $h/2e$ peak (dashed line) and
find excellent agreement.

\begin{figure}
  \begin{center}
    \includegraphics{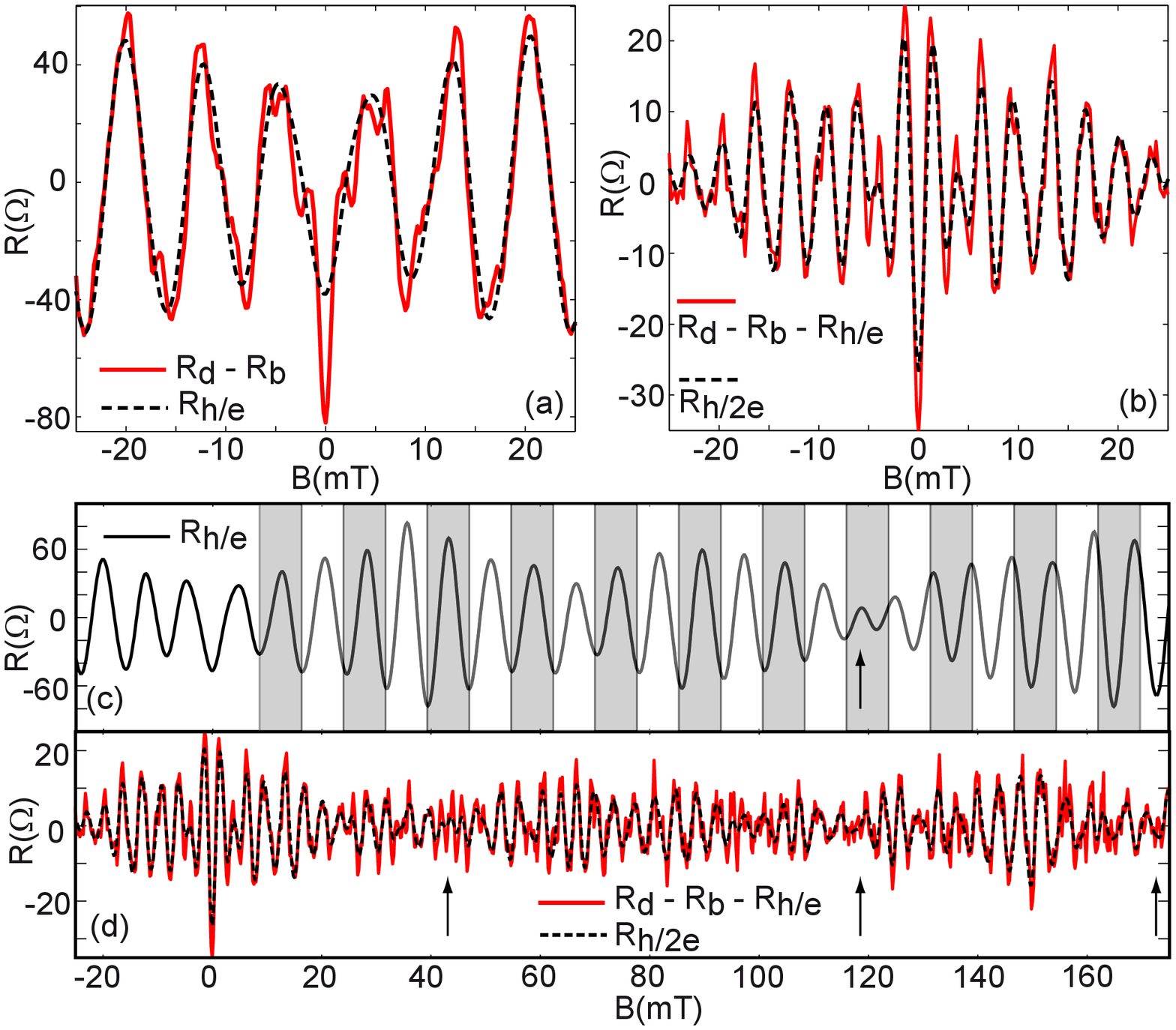}
    \caption{(color online) (a) Measured magnetoresistance of the ring after subtracting the low-frequency background, $R_d-R_b$ (full line, red online), together with the filtered $h/e$  oscillations $R_{h/e}$ (dashed line).
    (b) Difference $R_d-R_b-R_{h/e}$ (full line red online) together with the inverse Fourier transform of the $h/2e$ peak $R_{h/2e}$ (dashed line). (c) Beating in filtered $h/e$ oscillations. The width of the gray and white rectangles corresponds to the period of 7.7 mT. The arrow points to the beating node where a phase jump of $\pi$ occurs. (d) Beating in
filtered $h/2e$ oscillations with arrows indicating possible
nodes.}
    \label{fig2}
  \end{center}
\end{figure}

We further plot in Fig. 2(d) the difference $R_d-R_b-R_{h/e}$
(full line, red online), together with the filtered $h/2e$
oscillations $R_{h/2e}$ (dashed line) in a larger range of
magnetic fields. A beating-like behavior in the $h/2e$
oscillations is observed. Possible nodes develop around 40 mT, 115
mT, and 175 mT [arrows in Fig. 2(d)]. The appearance of these
unequally spaced nodes is in agreement with the complex split-peak
pattern in Fig. 1(e). In the plot of the filtered $h/e$
oscillations [Fig. 2(c)] we notice that only the node around 115
mT is common for both, $h/e$ and $h/2e$ oscillations, while the
other two nodes in the $h/2e$ oscillations correspond to maxima in
the beating of $h/e$ oscillations. This kind of aperiodic
modulation of the envelope function of the $h/2e$ oscillations,
rather than a regular beating, is predicted for the case of
diffusive rings in the presence of Berry's phase \cite{Engel00},
since the latter also changes with increasing external magnetic
field.

The evolution of the AB oscillations upon changing plunger gate
voltages $V_{pg1}$ and $V_{pg2}$ is explored in Fig. 3(a). Plunger
gate voltages are changed antisymmetrically: $V_{pg1}=-120 $mV
$-V$; $V_{pg2}=-120$ mV $+V$. Two distinct features are visible:
there is always a local minimum in the AB oscillations at $B=0$ T,
and the oscillations experience a phase jump by $\pi$ around
$V=27$ mV. In order to understand the origin of these two features
we analyze the filtered $h/e$ (not shown) and $h/2e$ oscillations
[Fig. 3(b)] as a function of $V$. The $h/e$ oscillations
experience a phase jump of $\pi$, while the $h/2e$ oscillations do
not [Fig. 3(b)]. We have explored this behavior in several other
gate configurations and always found the same result. The reason
for such a behavior is that the $h/e$ oscillations are sensitive
to the phase difference $\Delta\varphi=k_1l_1-k_2l_2$ between the
two arms, which can be changed by the plunger gates, while the AAS
$h/2e$ oscillations are not. 
We observe a resistance minimum at $B=0$ T for all gate
configurations [Fig. 3(a)], which is due to a minimum at $B=0$ T
in the $h/2e$ oscillations [Fig. 3(b)]. It indicates that time
reversed paths of the holes' spinors interfere destructively due
to strong SOI, in contrast to n-type GaAs systems where $h/2e$
oscillations produce a resistance maximum at $B=0$T \cite{Ihn03}.
This effect has the same origin as the weak anti-localization
(WAL) effect. However, the observed minimum is not caused by WAL
in the ring leads, since the WAL dip in bulk 2D samples has a much
smaller magnitude (less than $1\Omega$, \cite{GrbicThesis07}) than
the minimum at $B=0$T in the ring. The resistance minimum at
$B=0$T is a result of the destructive interference of the holes'
spins in the ring.

\begin{figure}
  \begin{center}
    \includegraphics{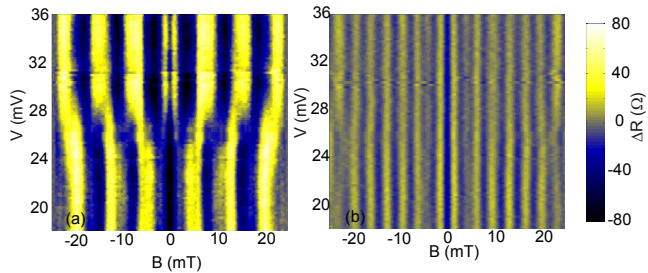}
    \caption{(color online) (a) Evolution of the AB oscillations upon changing the plunger gate
voltages $V_{pg1}=-120 $mV $-V$; $V_{pg2}=-120$ mV $+V$. (b)
Filtered $h/2e$ oscillations as a function of the plunger gate
voltages, showing the local minimum at $B=0$ T at all gate
voltages.}
    \label{fig3}
  \end{center}
\end{figure}

The adiabatic regime is reached when $\omega_B / \omega_{orbit} >>
1$, where $\omega_B=g \mu_B B_{tot} /2 \hbar$ is the spin
precession frequency around $\mathbf{B_{tot}}$, while
$\omega_{orbit}= v_F/r$ is the orbital frequency of the holes
around the ring in the ballistic regime. p-type GaAs systems have
strong SOI, and therefore large $B_{int}$, which, together with a
small $v_F$ (due to the large effective mass of the holes) makes
p-types systems very favorable for reaching the adiabatic regime
compared to other systems. Using the estimated value for $B_{int}$
of 0.25T and assuming a holes' $g$ factor of 2, we obtain
$\omega_B / \omega_{orbit}\approx 0.2-0.3$ for the measured range
of $B_{ext}$ up to 0.2T. Therefore the adiabatic regime is not
fully reached in our measurements.

There remains a pronounced discrepancy between the internal
magnetic field obtained from the beating of the SdH oscillations
of 7T (converting the corresponding energy scale
$\Delta_{SO}\approx0.8$ meV to a magnetic field via the Zeeman
splitting) and the field scale of 0.25 T obtained from the beating
of the AB oscillations. The latter evaluation is strictly valid in
the diffusive regime \cite{Engel00} while our sample is at the
crossover to the ballistic regime. It is also not clear how the
limited adiabaticity in our samples will influence these numbers.

In a straightforward picture one would expect that the node of the
beating in the $h/2e$ oscillations occurs at half the magnetic
field as the node in the $h/e$ oscillations since the accumulated
phase difference between the two spin species should be
proportional to the path length travelled in the ring. Within
experimental accuracy the data in Fig. 1 (d) and (e) suggests that
the splitting in the corresponding Fourier transforms is the same.

In conclusion, we have measured Aharonov-Bohm oscillations in a
ring defined on a 2D hole gas with strong spin-orbit interactions.
We observe a beating in the measured resistance which arises from
an interplay between the orbital Aharonov-Bohm and a spin-orbit
induced geometric phase. In addition we resolve $h/2e$
oscillations in the ring resistance, and find that they also show
a beating-like behavior, which produces a splitting of the $h/2e$
peak in the Fourier spectrum.
A resistance
minimum at $B=0$, present in all in-plane gate configurations,
demonstrates the destructive interference of the hole spins
propagating along time reversed paths.

We thank Daniel Loss and Yigal Meir for stimulating discussions. Financial
support from the Swiss National Science Foundation is gratefully
acknowledged.

\end{document}